\documentclass[twocolumn,shortnote]{jpsj2}

\def\runauthor{Akira {\sc Oguri}}

\title{ 
Transmission Coefficient as a 
Three-Point Retarded Function  
}

\author{Akira {\sc Oguri}}

\inst{
          Department of Material Science,
          Osaka City University, 
          Sumiyoshi-ku, Osaka 558-8585,
          Japan 
}

\recdate{\today}
 
\abst{
}

\kword{
transmission probability, 
interacting systems,
Kubo formula,
retarded function, real time
}

\begin{document}

\sloppy

\maketitle

Previously,
we have described a formulation for 
the transmission probability of small interacting systems 
connected to noninteracting leads.\cite{ao10}
Using the Kubo formula and a \'{E}liashberg theory 
\cite{Eliashberg} for the analytic continuation of vertex functions,
the dc conductance $g$ has been shown to be written in a Landauer-type form;  
 \begin{equation}
 g = (2e^2/h) \int {\rm d}\epsilon\, (- \partial f / \partial \epsilon) \, 
 {\cal T}(\epsilon) \;, 
 \end{equation}
where $f(\epsilon)$ is the Fermi function.  
In Ref.\ 1, we have provided the expression 
of the transmission probability ${\cal T}(\epsilon)$ 
in terms of a vertex function eq.\ (2.36). Also,  
we gave another expression, eq.\ (3.6) in Ref.\ 1, 
which uses a three-point correlation function. 
Both of the two expressions have been obtained after carrying out 
the analytic continuation for the Matsubara frequencies.

\begin{figure}[b]
\leavevmode 
\begin{center}
\begin{minipage}{0.8\linewidth}
\includegraphics[width=\linewidth, clip, 
trim = 4cm 18cm 3cm 2.5cm]{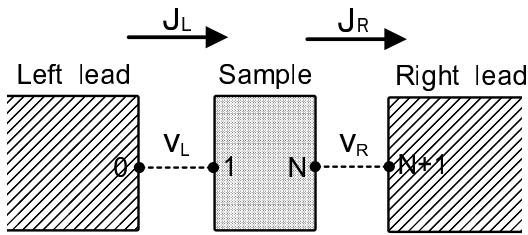}
\end{minipage}
\caption{ Schematic picture of the system.} 
\label{fig:single}
\end{center}
\end{figure}

The purpose of this short note is to show that
 ${\cal T}(\epsilon)$ can also be expressed in terms of
the retarded products of the three-point 
correlation function, and eq.\ (\ref{eq:time2}) is 
the main result of this report. 
Since the retarded product is defined in the real time, 
it gives us direct information about the relationship between  
the transmission probability and dynamic correlation functions.

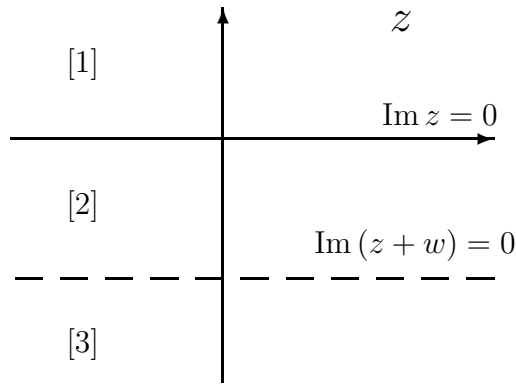
\begin{figure}[h]
\setlength{\unitlength}{0.6mm}
\begin{center}
\begin{picture}(108,85)(-54,-57) 
\thicklines

\put(-50,0){\vector(1,0){107}}
\put(-3,-54){\vector(0,1){83}}
\multiput(-49,-31)(10,0){11}{\line(1,0){6}}

\put(30,3){\makebox(0,0)[bl]{\large $\mbox{Im}\,z=0$}}
\put(15,-27){\makebox(0,0)[bl]{\large $\mbox{Im}\,(z+w)=0$}}

\put(34,24){\makebox(0,0)[bl]{\Large $z$}}

\put(-40,13){\makebox(0,0)[bl]{\large $[1]$}}
\put(-40,-18.5){\makebox(0,0)[bl]{\large $[2]$}}
\put(-40,-49){\makebox(0,0)[bl]{\large $[3]$}}

\end{picture}
\end{center}
\caption{Three analytic regions of 
$\Phi_{R;11}(z, z+w)$.
}
\label{fig:current_analytic}
\end{figure}

In the following, we will use the same notation 
with that in Ref.\ 1. 
A schematic picture of the system is illustrated in Fig.\ \ref{fig:single}.
The label $1$ ($N$) is assigned to the localized state 
at the interface on the left (right), 
and the label $0$ and $N+1$ are assigned to the site 
at the reservoir-side of the interface.
The inter-electron interaction is switched on only 
for the electrons inside the central region: 
the complete Hamiltonian ${\cal H}$ is given by eq.\ (2.1) of Ref.\ 1.
The transmission probability of this system can be written as 
(see eq.\ (3.6) of Ref.\ 1) 
\begin{eqnarray}
{\cal T}(\epsilon)  &=&     
  2\, \Gamma_L(\epsilon) \, \Phi_{R;11}^{[2]}(\epsilon, \epsilon) 
\;,
\label{eq:T_eff2}
\end{eqnarray}
where $\Gamma_L(\epsilon)=\pi  \rho_L^{\phantom{0}}(\epsilon)  v_L^2$, 
$\rho_L^{\phantom{0}}(\epsilon)$ is 
the local density of states at the site \lq\lq$0$" in the left lead, 
and $v_L$ is the mixing matrix element 
which connects the sample and the left lead.
The three-point function 
$\Phi_{R;11}^{[2]}(\epsilon, \epsilon +\omega)$ 
has been introduced using the imaginary-time formulation, 
\begin{align}
&\Phi_{R;11}(\tau ; \tau_1, \tau_2) \ = \
\left \langle  T_{\tau} \, J_{R}(\tau)\,
 c_{1\sigma}^{\phantom{\dagger}}(\tau_1) \, c_{1\sigma}^{\dagger}(\tau_2)       
 \right \rangle  \;,
\label{eq:3point_Matsubara} 
\\ 
&
\int_0^{\beta} 
\text{d}\tau\, \text{d}\tau_1\, \text{d}\tau_2\, 
\text{e}^{\text{i}\nu \tau} 
\text{e}^{\text{i}\varepsilon  \tau_1} 
\text{e}^{-\text{i}\varepsilon ' \tau_2}
\ \Phi_{R;11}(\tau ; \tau_1, \tau_2) 
\nonumber \\
&
\!\!\!\!
 = \  \beta\, 
    \delta_{\varepsilon+\nu,\varepsilon'}\,
    \Phi_{R;11}(\text{i}\varepsilon , \text{i}\varepsilon +\text{i}\nu)\; ,
\end{align}
where  $J_R \equiv 
 {\rm i}\,   \sum_{\sigma}
         v_{R}^{\phantom{\dagger}}
  \left(\, 
           c^{\dagger}_{N+1 \sigma} 
           c^{\phantom{\dagger}}_{N  \sigma} 
        -  c^{\dagger}_{N \sigma} 
           c^{\phantom{\dagger}}_{N+1  \sigma} 
\, \right) 
$ is the current flowing through the right interface, 
$J_R(\tau) \equiv e^{\tau {\cal H}} J_R e^{-\tau {\cal H}}$,
and $c^{\dagger}_{1 \sigma}$ creates an electron with spin $\sigma$ at 
the site \lq\lq$1$" at the left interface.  
As a function of complex variables,
 $\Phi_{R;11}(z , z+w)$ has singularities along the lines 
 $\mbox{Im}\,(z)=0$ and $\mbox{Im}\,(z+w)=0$,  
 which can be seen in eq.\ (\ref{eq:Lehmann_1}).
These two singularities divide the complex $z$ plane into three regions 
(see Fig.\ \ref{fig:current_analytic}),  
in each of which $\Phi_{R;11}(z,z+w)$ corresponds to 
the analytic function 
\begin{equation}
\left\{
\begin{array}{l}
\Phi_{R;11}^{[1]}(\epsilon, \epsilon+\omega) \ = \ 
\Phi_{R;11}(\epsilon +\text{i}0^+, \epsilon + \omega +\text{i}0^+)
\\
\Phi_{R;11}^{[2]}(\epsilon, \epsilon+\omega) \ = \ 
\Phi_{R;11}(\epsilon -\text{i}0^+, \epsilon + \omega + \text{i}0^+)
\\
\Phi_{R;11}^{[3]}(\epsilon, \epsilon+\omega) \ = \ 
\Phi_{R;11}(\epsilon -\text{i}0^+, \epsilon + \omega -\text{i}0^+)
\end{array} \right. 
\label{eq:analytic_continuation}
\;.
\end{equation}

Therefore it is the analytic continuation in the region [2],    
i.e., $\Phi_{R;11}^{[2]}(\epsilon, \epsilon +\omega)$, 
that determines the transmission probability through 
eq.\ (\ref{eq:T_eff2}).
The aim of this report is to present another approach 
to get this function staring from the real time 
without carrying out the analytic continuation.
To this end, we consider the Lehmann representation for 
$\Phi_{R;11}(\text{i}\varepsilon , \text{i}\varepsilon +\text{i}\nu)$.
Inserting a complete set of the eigenstates satisfying 
${\cal H}|n\rangle = E_n |n\rangle$  into eq.\ (\ref{eq:3point_Matsubara}), 
we have   
\begin{align}
& \Phi_{R;11}(\text{i}\varepsilon , \text{i}\varepsilon +\text{i}\nu)
 \nonumber \\
 & =
\ {1 \over Z} \sum_{lmn} \,
\langle l|c_{1\sigma}^{\dagger}|m\rangle
\langle m|J_R|n\rangle
\langle n|c_{1\sigma}^{\phantom{\dagger}}|l\rangle 
\nonumber\\
& \ \ 
\times
\biggl[\,
{ \text{e}^{-\beta E_m} \over 
  (\text{i}\varepsilon +\text{i}\nu + E_m-E_l)(\text{i}\nu +E_m -E_n)} 
\nonumber \\  
& \ \ \    
-\,{ \text{e}^{-\beta E_l} \over 
  (\text{i}\varepsilon  + E_n-E_l)
  (\text{i}\varepsilon +\text{i}\nu +E_m -E_l)} 
\nonumber \\  
& \ \ \   
-\,{ \text{e}^{-\beta E_n} \over 
  (\text{i}\nu + E_m-E_n)(\text{i}\varepsilon  +E_n -E_l)} 
\,\biggr] 
\nonumber \\  
& 
\nonumber \\  
& \ \ 
 +     
 {1 \over Z} \sum_{lmn} \,
\langle l|c_{1\sigma}^{\phantom{\dagger}}|n\rangle
\langle n|J_R|m\rangle
\langle m|c_{1\sigma}^{\dagger}|l\rangle 
\nonumber\\
& \ \ \ \ \times
\biggl[\,
{ \text{e}^{-\beta E_n} \over 
  (\text{i}\varepsilon  + E_l-E_n)(\text{i}\nu +E_n -E_m)} 
\nonumber \\  
& \ \ \    
+\,{ \text{e}^{-\beta E_l} \over 
  (\text{i}\varepsilon  + E_l-E_n)
  (\text{i}\varepsilon +\text{i}\nu +E_l -E_m)} 
\nonumber \\  
& \ \ \   
-\,{ \text{e}^{-\beta E_m} \over 
  (\text{i}\varepsilon  +\text{i}\nu+ E_l-E_m)(\text{i}\nu +E_n -E_m)} 
\,\biggr] , 
\label{eq:Lehmann_1}
\end{align}
where $Z = \mbox{Tr}\, \text{e}^{-\beta {\cal H}}$. 
From eq.\ (\ref{eq:Lehmann_1}),  
we can also obtain the Lehmann representation 
of $\Phi_{R;11}^{[k]}(\epsilon, \epsilon+\omega)$ for 
$k=1,\,2,\,3$ by replacing the imaginary frequencies 
$\text{i}\varepsilon$ and $\text{i}\nu$ with  
the real ones $\epsilon$ and $\omega$, respectively, 
taking the infinitesimal imaginary parts summarized in the right-hand side 
of eq.\ (\ref{eq:analytic_continuation}) into account. 
Then, it is straight forward to show that the same analytic functions 
can be derived form the real-time functions 
defined by 
\begin{align}
&  \Phi_{R;11}^{[1]}(t ; t_1, t_2) 
\nonumber \\
&  
 = \ 
\ \theta(t-t_1)\,
\theta(t_1-t_2)\,
\left \langle 
\left[\,
\left\{ c_{1\sigma}^{\phantom{\dagger}}(t_1) 
\,,\, c_{1\sigma}^{\dagger}(t_2)\right\} 
\,, J_R(t) \right]\,
 \right \rangle  
 \nonumber \\
&  \ \ \  + \, \theta(t_1-t)\,
\theta(t-t_2)\,
\left \langle 
\left\{ c_{1\sigma}^{\phantom{\dagger}}(t_1) \,,\, 
\left[c_{1\sigma}^{\dagger}(t_2)\,,\, J_R(t) \right] \right\}
\, \right \rangle  ,
 \nonumber \\
\label{eq:time1}
\\
&  \Phi_{R;11}^{[2]}(t ; t_1, t_2) 
\nonumber \\
&  
= \ 
\ \theta(t-t_1)\,
\theta(t_1-t_2)\,
\left \langle 
\left\{ c_{1\sigma}^{\dagger}(t_2) \,,\, 
\left[c_{1\sigma}^{\phantom{\dagger}}(t_1)\,,\, J_R(t) \right] \right\}
\, \right \rangle  
\nonumber \\
&  \ \ \  - \, \theta(t-t_2)\,
\theta(t_2-t_1)\,
\left \langle 
\left\{ c_{1\sigma}^{\phantom{\dagger}}(t_1) \,,\, 
\left[c_{1\sigma}^{\dagger}(t_2)\,,\, J_R(t) \right] \right\}
\, \right \rangle ,  
 \nonumber \\
\label{eq:time2}
\\
&  \Phi_{R;11}^{[3]}(t ; t_1, t_2) 
\nonumber \\
&  
 = \ 
-\, \theta(t-t_2)\,
\theta(t_2-t_1)\,
\left \langle 
\left[\,
\left\{ c_{1\sigma}^{\phantom{\dagger}}(t_1) 
\,,\, c_{1\sigma}^{\dagger}(t_2)\right\} 
\,, J_R(t) \right]\,
 \right \rangle  
 \nonumber \\
&  \ \ \  - \, \theta(t_2-t)\,
\theta(t-t_1)\,
\left \langle 
\left\{ c_{1\sigma}^{\dagger}(t_2) \,,\, 
\left[c_{1\sigma}^{\phantom{\dagger}}(t_1)\,,\, J_R(t) \right] \right\}
\, \right \rangle  , 
\nonumber \\
\label{eq:time3}
\end{align}
through the Fourier transform   
\begin{align}
& \int_{-\infty}^{\infty} 
\text{d}t\, \text{d}t_1\, \text{d}t_2\, 
\text{e}^{\text{i}\omega t} 
\text{e}^{\text{i}\epsilon  t_1} 
\text{e}^{-\text{i}\epsilon' t_2}\, 
\Phi_{R;11}^{[k]}(t ; t_1, t_2) 
\nonumber \\
& 
 = \ 
     2\pi\, \delta(\epsilon+\omega-\epsilon')\, 
    \Phi_{R;11}^{[k]}(\epsilon, \epsilon + \omega) \;. 
\label{eq:Fourier_time}
\end{align}
Here 
$J_R(t) \equiv e^{\text{i} {\cal H} t} J_R e^{- \text{i} {\cal H}t}$
and $\theta(t)$ is the step function.
The two types of the brackets denote 
the commutator $[A,B] \equiv AB-BA$, 
and the anticommutator $\{A,B\}\equiv AB+BA$. 
One can confirm 
the above statement about eqs.\ (\ref{eq:time1})--(\ref{eq:time3}) 
by carrying out the integration in eq.\ (\ref{eq:Fourier_time}) 
explicitly. For instance,  the Fourier transform of a function 
\begin{equation}
F(t ; t_1, t_2)  = 
\theta(t-t_1)\,
\theta(t_1-t_2)
\left \langle J_R(t)\,
 c_{1\sigma}^{\phantom{\dagger}}(t_1) \, c_{1\sigma}^{\dagger}(t_2)       
 \right \rangle 
\end{equation}
can be calculated as 
\begin{align}
&  F(\epsilon,\epsilon+\omega) 
\nonumber \\ 
&=
 {-1 \over Z} \sum_{lmn} \,
{
\text{e}^{-\beta E_m} \, 
\langle l|c_{1\sigma}^{\dagger}|m\rangle
\langle m|J_R|n\rangle
\langle n|c_{1\sigma}^{\phantom{\dagger}}|l\rangle 
\over
  (\epsilon + \omega + E_m-E_l +\text{i}0^+)(\omega +E_m -E_n+\text{i}0^+)
  } 
  \;.
\nonumber \\
\end{align}
Among the three real-time functions eqs.\ (\ref{eq:time1})--(\ref{eq:time3}),
only $\Phi_{R;11}^{[2]}(t ; t_1, t_2)$ in eq.\ (\ref{eq:time2}) is 
relating to the transmission probability: 
eqs.\ (\ref{eq:time1}) and (\ref{eq:time3}) are provided  
for comparison.
Note that,  
in both of the two averages in the right-hand side of eq.\ (\ref{eq:time2}), 
the commutator for one fermion operator and current $J_R$ is situated inside 
the anticommutator for another fermion operator.

In conclusion  the three-point function 
$\Phi_{R;11}^{[2]}(\epsilon, \epsilon + \omega)$, 
which determines the transmission probability eq.\ (\ref{eq:T_eff2}), 
can be described as the Fourier transform of 
the real-time correlation function $\Phi_{R;11}^{[2]}(t ; t_1, t_2)$ introduced in eq.\ (\ref{eq:time2}). 
This real-time formulation seems to be applicable to 
nonperturbative approaches to the transmission probability 
${\cal T}(\epsilon)$ of correlated electron systems. 


This work is supported by the Grant-in-Aid 
for Scientific Research from the Ministry of Education, 
Culture, Sports, Science and Technology,
Japan.


\end{document}